\documentclass[aps,prb,twocolumn,groupedaddress,showpacs,superscriptaddress,amssymb,amsmath]{revtex4-1}
\usepackage{graphicx}
\usepackage{tabularx}
\usepackage{color}
\usepackage{amsmath}
\usepackage{comment}
\newcommand{\be}{\begin{equation}}
\newcommand{\ee}{\end{equation}}
\newcommand{\bea}{\begin{eqnarray}}
\newcommand{\eea}{\end{eqnarray}}
\usepackage{dcolumn}
\usepackage{hyperref}
\usepackage{bm}
\usepackage{epsf}
\usepackage{subfigure}
\usepackage{epstopdf}%
\usepackage{amsfonts}

\begin{document}

\title{On the validity of the thermodynamic uncertainty relation in quantum systems}

\author{Bijay Kumar Agarwalla}
\affiliation{Department of Physics, Dr. Homi Bhabha Road, Indian Institute of Science Education and Research, Pune, India 411008}

\author{Dvira Segal}
\affiliation{Chemical Physics Theory Group, Department of Chemistry, and
Centre for Quantum Information and Quantum Control,
University of Toronto, 80 Saint George St., Toronto, Ontario, Canada M5S 3H6}

\date{\today}

\begin{abstract}
We examine the so-called thermodynamic uncertainty relation (TUR), a cost-precision trade-off relationship
in transport systems. 
Based on the fluctuation symmetry, we derive a condition on the validity of the TUR for general 
nonequilibrium (classical and quantum) systems.
We find that the first non-zero contribution to the TUR beyond equilibrium, 
given in terms of nonlinear transport coefficients, can be positive or negative,
thus affirming or violating the TUR depending on the details of the system.
We exemplify our results for noninteracting quantum systems by deriving the
thermodynamic uncertainty relation in the language of the transmission function.
We demonstrate that quantum coherent systems that do not follow a population
Markovian master equation, e.g. by supporting high order tunneling processes or relying on coherences, violate the TUR.


\end{abstract}

\maketitle 

\section{Introduction}
\label{Sec-intro}

A thermodynamic uncertainty relation (TUR) describing a trade-off between 
entropy production (cost) and precision (noise)
was recently derived for classical markovian systems operating at steady state
\cite{TUR1,TUR2,TUR3,TUR4,TUR5}, with generalizations to finite-time  \cite{TUR6},
time-discrete and driven Markov chains \cite{TUR7}.
The thermodynamic uncertainty relation has generated significant research work
directed at understanding its ramifications to dissipative systems such as
biochemical motors and heat engines \cite{TURrev} and probing its validity 
in the classical regime \cite{lanj1,lanj2} and beyond  \cite{TURQ,TURcoh}.

For a two-terminal single-affinity system, the TUR connects the steady state averaged current 
$\langle j\rangle$, its variance $\langle \langle j^2\rangle\rangle$, and the entropy production 
rate $\sigma$ in a nonequilibrium process as
\bea
\frac{\langle \langle j^2\rangle \rangle}{  \langle j\rangle^2} \frac{\sigma }{ k_B} \geq 2.
\label{eq:TUR0}
\eea
This relation reduces to an equality in linear response:
the current and the entropy production rate are linear in the thermodynamic affinity $A$, 
$\langle j \rangle =  G A T$, ($T$ is the temperature), $\sigma=\langle j \rangle A$, 
and the noise satisfies the fluctuation-dissipation relation with the linear 
transport coefficient $G$, $\langle \langle j^2 \rangle \rangle = 2\,G\,k_B\,T$.
Away from equilibrium, Eq. (\ref{eq:TUR0}) points to 
a fundamental trade-off between precision and dissipation: 
A precise process with little noise is realized with a high thermodynamic (entropic) cost.
We refer to systems that obey this inequality as ``satisfying the TUR". 
TUR violations correspond to situations in which the left hand side of Eq. 
(\ref{eq:TUR0}) is smaller than 2.

Several interesting questions immediately come to mind when inspecting the TUR.
Does it hold for other systems beyond Markov processes? 
Away from equilibrium, can we derive the TUR from fundamental principles, essentially from 
the fluctuation symmetry \cite{fluct1,fluct2,fluct3,fluct4,fluct5}? 
What is the role of quantum effects in validating or violating the inequality \cite{TURQ}? 
Recent studies addressed the potential of quantum coherences to reduce fluctuations in 
quantum heat engines \cite{TURcoh}. 

The objective of our work is to understand the validity/invalidity of
the TUR from fundamental principles beyond specific examples. In particular:
(i) We derive the TUR from the steady state nonequilibrium fluctuation symmetry and 
achieve conditions for validating or violating it. (ii) We study quantum systems
that do not satisfy Markovian dynamics and rationalize 
violations of the TUR in certain regimes of operation.

We emphasize that multi-affinity systems such as thermoelectric junctions 
go beyond the lower bound of 2 even close to equilibrium \cite{BrandLR}, while a single-affinity
system shows an equality in linear response. 
In this work, we focus our attention on the latter situation, where a
far-from-equilibrium condition is necessary for departing from an equality in Eq. (\ref{eq:TUR0}).

The paper is organized as follows. In Sec. \ref{Sec-FR}, we derive the TUR from the fundamental 
fluctuation symmetry. We analyze fermionic quantum transport junctions in Sec. \ref{Sec-ferm}
by deriving a closed-form condition on the TUR.
We study examples in Sec. \ref{Sec-examples} and conclude in Sec. \ref{Sec-summ}.

\section{Derivation of the TUR from the fluctuation relation}
\label{Sec-FR}

The TUR was derived in Refs. \cite{TUR1,TUR2,TUR3} for classical Markov processes. Here, 
we arrive at a general connection between the current and its fluctuations for arbitrary systems, 
classical or quantum, based on the fluctuation symmetry.
The setup that we have in mind is a junction, where a finite size system is 
sandwiched between two fermionic leads with a bias voltage $V$.
The steady state fluctuation symmetry \cite{fluct1,fluct2,fluct3,fluct4,fluct5} relates
the probability for transferring $n$ carriers from high to low voltage over the time interval $t$, 
$P_t(n)$, to the probability of transferring charges against the applied voltage, $P_t(-n)$ (assuming $e=1$),
\be
 \ln \Big[\frac{P_t(n)}{P_t(-n)}\Big]= \beta \,V n,
\label{eq:FS}
\ee
where $\beta=1/k_B T$ is the inverse temperature. 
We define the characteristic function, ${\cal Z}(\alpha)\equiv \langle e^{i\alpha n}\rangle=  
\sum_{n=-\infty}^{\infty}P_t(n)e^{i\alpha n}$,
%
and the long time limit of the cumulant generating function (CGF) as
$\chi(\alpha)= \lim_{t \to \infty} \frac{1}{t}\ln {\cal Z(\alpha)}$, or equivalently 
\bea
\chi(\alpha)=\lim_{t \to \infty} \frac{1}{t} \sum_{p=1}^{\infty} \langle \langle n^p\rangle \rangle \frac{(i\alpha)^p}{p!}.
\eea
The steady state fluctuation relation (\ref{eq:FS}) dictates the symmetry
$\chi(\alpha) = \chi(-\alpha + i \beta V)$.
The fluctuation symmetry ensures relations between transport coefficients, specifically \cite{fluct3} 
\begin{eqnarray}
S_0&=& 2 \,k_B \,T \,G_1, 
\nonumber\\
S_1&=& k_B \,T \,G_2.
\label{eq:S01}
\end{eqnarray}
The first equation is the Johnson-Nyquist (fluctuation-dissipation) relation. 
The second equation uncovers a universal
relation in the nonlinear transport regime, beyond the Onsager relation.
Here, the steady state charge current and its associated fluctuations (noise) 
are expanded in order of applied voltage,
\begin{eqnarray}
\langle j \rangle &=& G_1 V + \frac{1}{2!} G_2 V^2 + \frac{1}{3!} G_3 V^3 \cdots 
\nonumber \\
\langle \langle j^2 \rangle \rangle &=& S_0 + S_1 V + \frac{1}{2!} S_2 V^2 + \frac{1}{3!} S_3 V^3 \cdots .
\end{eqnarray}
Here, $G_1, G_2, G_3 \cdots$ are the linear and non-linear transport coefficients. Similarly, 
$S_0, S_1, S_2 \cdots$ are the equilibrium and non-equilibrium noise terms.
The associated entropy production is the joule's heating, $\sigma= V \langle j \rangle/T$.  

We are now ready to put these relations together, and we find that
\bea
\beta V \frac{\langle \langle j^2 \rangle \rangle}{\langle j \rangle} = 2 + \frac{V^2}{G_1} C_{\rm neq} + {\mathcal O}(V^4) + \cdots 
\label{eq:TUR}
\eea
%
where we define the function $C_{\rm neq}$ as
\bea
C_{\rm neq} = \frac{\beta}{6} \Big[ 3 S_2 - 2 k_B T G_3\Big].
\label{eq:f2}
\eea
%
%
%
Note that an $\mathcal{O}(V)$ term does not appear in the above expression 
precisely due to the relations (\ref{eq:S01}).  

Equations (\ref{eq:TUR})-(\ref{eq:f2}) are central results of this work: a combination of nonlinear transport
coefficients determine the validity of the TUR.
In the $\mathcal{O}(V^2)$, the TUR is satisfied if 
$C_{\rm neq}\geq 0$. It is violated once $C_{\rm neq}<0$.
Eq. (\ref{eq:TUR}) holds for arbitrary classical or quantum junctions. The only underlying
requirement behind it are relationships between transport coefficients, in and beyond linear response.
One can further extend this result for systems under a magnetic field and for bosonic systems. 
A nontrivial consequence of our work is that for Markov processes,
high order transport coefficients satisfy an inequality, $S_2\geq \frac{2}{3}k_BTG_3$.

\vspace{3mm}
\section{Noninteracting charge transport: Formula for the TUR violation} 
\label{Sec-ferm}


We focus here on a generic noninteracting quantum charge transport problem and
study the validity ($C_{\rm neq} \geq0$) and breakdown ($C_{\rm neq}<0$) of the TUR.
We consider a tight-binding chain connected to two fermionic leads that are 
maintained at different chemical potentials but at the same temperature. 
The steady state cumulant generating function associated with the 
integrated charge current was first derived by Levitov-Lesovik \cite{Levitov1, Levitov2,Klich}.
It was later extended to finite-size systems \cite{FCS1,FCS2}, and written
following the Keldysh nonequilibrium Green's function formalism \cite{fluct1,bijay-PRE-fcs}. 
%
It is given by
\bea
\chi(\alpha) &=& \int_{-\infty}^{\infty} \frac{dE}{2 \pi \hbar} \ln \Big[ 1 + {\cal T}(E) \big[ f_L (E) (1- f_R(E)) (e^{i \alpha}-1) \nonumber \\
&+& f_R(E) (1-f_L(E))(e^{-i \alpha}-1) \big]\Big].
\label{eq:CGF}
\eea
Here, ${\cal T}(E)$ is the transmission function for charge transport at energy $E$. It
can be calculated from the retarded and advanced Green's function of the system and from its self energy matrix 
\cite{Nitzan,diventra}.
The transmission function is restricted to $0 \leq {\cal T}(E) \leq 1$. 
$\alpha$ is a counting parameter for charge transfer. 
$f_{\nu}(E) = \big[e^{\beta (E-\mu_{\nu})} + 1\big]^{-1}$ is the Fermi distribution function for the two
metal electrodes $\nu=L,R$.
As mentioned above, 
the CGF satisfies the steady state fluctuation symmetry, $\chi(\alpha) = \chi(-\alpha + i \beta V)$, 
where $V = \mu_L -\mu_R$ \cite{Gallavotti}. 
From the CGF, one can generate all the cumulants. 
In particular, to interrogate the TUR we need to 
focus on the current and its fluctuations, given as 
$\langle j \rangle = \frac{\partial \chi }{\partial (i\alpha)}|_{\alpha=0} $, 
$\langle\langle j^2 \rangle\rangle = \frac{\partial^2 \chi }{\partial (i\alpha)^2}|_{\alpha=0} $.  
Working out Eq. (\ref{eq:CGF}) we get
%
\begin{widetext}
\bea
\langle j \rangle &=& \int_{-\infty}^{\infty} \frac{dE}{2 \pi \hbar} {\cal T}(E) \big[f_L (E) - f_R(E)\big],
\nonumber \\
\langle \langle j^2 \rangle \rangle &=&
\int_{-\infty}^{\infty} \frac{dE}{2 \pi \hbar} \Big\{{\cal T}(E)  \big[f_L (E) (1 -f_L(E))  + f_R(E)(1-f_R(E))\big] 
+ {\cal T}(E) (1-{\cal T}(E)) (f_L(E)-f_R(E))^2 \Big\}.
\label{eq:levitov}
\eea
\end{widetext}
To test Eq.~(\ref{eq:TUR}), we
Taylor expand the current and its noise in orders of $V$ around equilibrium, $\mu_L =\mu + V/2$ and
$\mu_R = \mu-V/2$, so as to receive the transport coefficients $G_3$ and $S_2$, given by
\begin{eqnarray}
G_3 &=& \frac{1}{4} \int_{-\infty}^{\infty} \frac{dE}{2\pi \hbar} {\cal T}(E) \frac{\partial^3 f(E) }{\partial \mu^3},
\nonumber \\
S_2 &=& 2 \,k_B T \,G_3 \!+ 2 \int_{-\infty}^{\infty} \!\frac{dE}{2\pi \hbar} {\cal T}(E) (1\!-\!{\cal T}(E)) \left(\frac{\partial f(E) }{\partial \mu}\right)^2.
\nonumber\\
\end{eqnarray} 
Below we also make use of the linear conductance,
\bea
G_1=\int_{-\infty}^{\infty} \frac{dE}{2\pi \hbar} {\cal T}(E) \frac{\partial f(E)}{\partial \mu}.
\eea
Substituting these coefficients into Eq. (\ref{eq:f2}), we gather 
\bea
&&\beta V \frac{\langle \langle j^2 \rangle \rangle}{\langle j \rangle} = 2  
 \nonumber\\
 &&+ \frac{\beta^2 V^2}{6 G_1}\! \int_{-\infty}^{\infty} \frac{dE}{2\pi \hbar}\, {\cal T}(E) \frac{\partial f(E)}{\partial \mu}\Big[1 -6\,{\cal T}(E) f(E) (1-f(E))\Big]
 \nonumber\\
 &&+ {\cal O} (V^4).
\label{eq:TURferm}
\eea  
Here, $f(E)$ is the equilibrium Fermi function. 
Following Eq.~(\ref{eq:TUR}) we identify $C_{\rm neq}$ by
\be
{C_{\rm neq}} \equiv \frac{\beta^2}{6}\!\int_{-\infty}^{\infty} \!\frac{dE}{2\pi \hbar} \, {\cal T}(E) \frac{\partial f(E) }{\partial \mu}\Big[1\!-\!6\,{\cal T}(E) f(E) (1-f(E))\Big].
\label{eq:F}
\ee
%
This function can switch sign depending on the structure of the transmission function as well as the 
contribution coming from the Fermi distribution. 
The validity of the TUR thus delicately depends on the properties of the system, i.e., its
energetics and its coupling to the leads, hidden within the transmission function,
as well as external conditions, i.e. the temperature.  

Equation (\ref{eq:F}) is a central result of this work. Evaluating it gives a direct measure
for TUR violation within the $V^2$ order. 
We now discuss several limiting cases of this formula.

{\it Low transmission, ${\cal T}(E) \ll1$}. In this limit,
the quadratic $[{\cal T}(E)]^2$ expression is ignored relative to ${\cal T}(E)$,
and we get (up to $V^2$),
\be
\beta V \frac{\langle \langle j^2 \rangle \rangle}{\langle j \rangle} = 2 + \frac{\beta^2 V^2}{6}.
\label{eq:low-trans}
\ee
Since ${C_{\rm neq}}=\frac{\beta^2 G_1}{6}>0$, the TUR is satisfied. 

{\it Constant Transmission.}
If the transmission is a constant independent of energy, ${\cal T}(E) =\tau$, 
one can show that the TUR is always satisfied. We note
that $\beta f(1-f)=\frac{\partial f}{\partial \mu}$ and $G_1=\frac{1}{2\pi\hbar}\tau$.
We then perform the integration in Eq. (\ref{eq:F}) using  
$\int dE \left( \frac{\partial f  }{\partial \mu}\right) =1$ and
$\frac{6}{\beta}\int dE \left( \frac{\partial f  }{\partial \mu}\right)^2 =1$, and obtain
\bea
\beta V \frac{\langle \langle j^2 \rangle \rangle}{\langle j \rangle}  = 2 + \frac{\beta^2 V^2}{12 \pi \hbar G_1} \tau ( 1-\tau) \geq 2.
\eea
For low transmission values, we recover Eq.~(\ref{eq:low-trans}).
For a perfect conductor or when approaching zero transmission, the TUR touches the lower (equilibrium) bound. 

{\it Resonance tunneling condition.} 
We can greatly simplify Eq. (\ref{eq:F}) when the metal-system hybridization is weak.
In this case, we assume that the transmission function is sharply peaked about a certain frequency 
$\epsilon_d$, which is set close to the Fermi energy,
 while the derivative of the Fermi function is relatively broad ($k_BT > \Gamma$). 
In this case, the principal contribution to the integral in Eq. (\ref{eq:F}) 
comes from the region near the resonance frequency $\epsilon_d$. We 
can therefore replace the derivative of the Fermi function by a constant, and simplify Eq. (\ref{eq:F}),
\be
{C}_{\rm neq} = \frac{\beta^3}{6} f(\epsilon_d) (1-f(\epsilon_d)) \Big[ {\cal T}_1 - 6 f(\epsilon_d) (1-f(\epsilon_d)) {\cal T}_2 \Big],
\ee
%
where we define the integrals
\bea {\cal T}_n \equiv \int_{-\infty}^{\infty}  \frac{dE}{2\pi \hbar} \, {\cal T}^n (E).
\label{eq:Tn}
\eea
Under this weak coupling approximation, the violation of TUR ($C_{\rm neq}<0$) translates into the inequality
\be
\frac{{\cal T}_1}{{\cal T}_2} < 6 f(\epsilon_d) (1-f(\epsilon_d)).
\ee
%
Since $0<f(1-f)<\frac{1}{4}$, 
the violation condition reduces to 
\be
\frac{{\cal T}_2}{{\cal T}_1} > \frac{2}{3}.
\label{eq:cond-trans}
\ee
This inequality is another important result of our work. 
It should be pointed out that this condition for the violation of the TUR
likewise holds for systems with multiple resonances $\epsilon_n$, 
as long as these resonances are sharp and sufficiently close,  $\epsilon_n, \Gamma_n < k_B T$,
such that they are all positioned
under the (approximately constant) envelope of $\frac{\partial f}{\partial \mu}$.
Here, $\Gamma$ is the characteristic  width of the resonances.

\begin{figure} 
\includegraphics[scale=1.8]{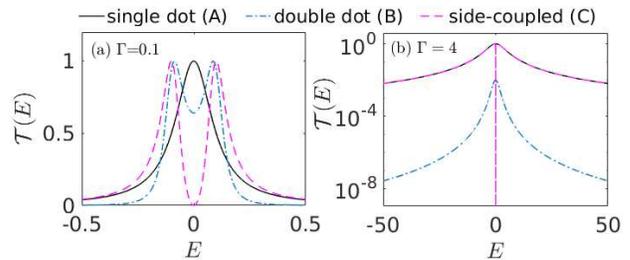}
\caption{
Transmission functions for models A, B, and C
at (a) weak and (b) strong couplings to the metal leads.
Model A, single dot junction: $\epsilon_d=0$;
Model B,  serial double dot junction: $\epsilon_{L,R}=0$ and $\Omega=0.1$.
Model C, side-coupled double dot junction: $\epsilon_{1,2}=0$ and $\Omega=0.1$.
}
\label{FigT}
\end{figure}


\section{Examples}
\label{Sec-examples}
We now examine three central charge transport models: a single dot resonant transmission model (A),
a serial double dot junction (B), and a side-coupled double dot model (C). 
All three models show violations of the TUR in certain regimes, as we demonstrate and rationalize in this Section. 

The transmission functions of the
three models are described in the Appendix and displayed in Fig. \ref{FigT}. 
While model A depicts a simple resonant-Lorentzian structure, models B and C show quantum interference effects.
In particular, in model B, when $\Omega$ is small, the system includes two quasi-degenerate levels
(in the energy basis), and the transmission is
suppressed at large $\Gamma$ when both levels contribute.
The side-coupled model displays a node at $E=0$
due to a destructive interference effect. In model C at large $\Gamma$, the transmission peak still 
reaches the unit value close to $E=0$, unlike model B.
We note that the double dot models B and C were recently analyzed in great details 
for demonstrating destructive quantum interference effects in molecular electronic conduction
 \cite{interf1,interf2,interf3,interf4}. 

Junctions A and B were recently studied in Ref. \cite{TURcoh} based on the Levitov-Lesovik formula, as well
as with a local Lindblad equation, manifesting TUR violations in different regimes. 
Our analysis in Secs. \ref{Sec-FR} and \ref{Sec-ferm} lays out the theoretical groundwork for 
these curious observations, and grants us with a fundamental understanding over TUR violations.

In calculations below we position the equilibrium Fermi energy at $\mu=0$. 
Furthermore, for simplicity, we consider only symmetric junctions with identical hybridization energies,  
$\Gamma=\Gamma_{L,R}$. 
In simulations we compare analytical results, referring to the second order in $V$ formula,
Eq.~(\ref{eq:TURferm}), to exact calculations using Eq.~(\ref{eq:levitov}).
For convenience, we also define the function
\bea
{\cal F} \equiv \beta V \frac{\langle \langle j^2\rangle \rangle}{\langle j \rangle} -2,
\eea
where the breakdown of the TUR corresponds to ${\cal F}<0$.

\begin{figure} [htbp]
\begin{center}
\includegraphics[scale=0.6]{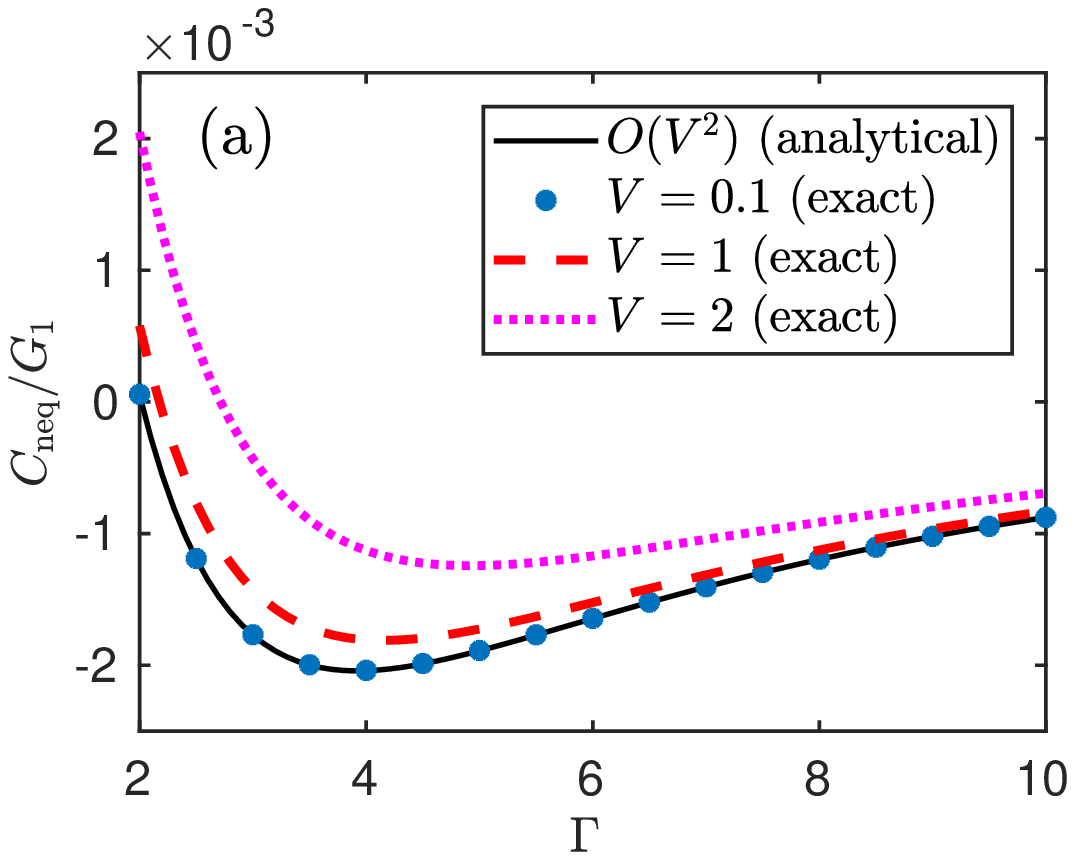}
\includegraphics[scale=0.6]{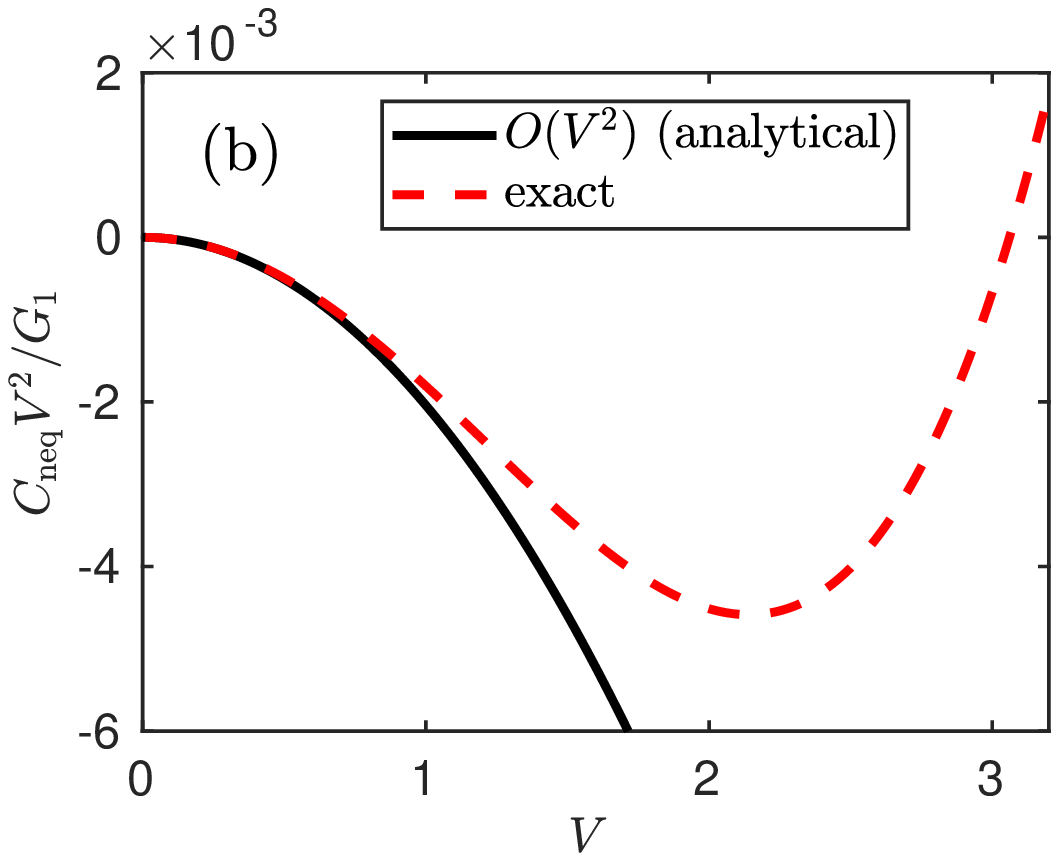}
\caption{
Single quantum dot: 
(a) Violation of the TUR as a function of metal-molecule coupling $\Gamma$ compared to
exact calculations for ${\cal F}/V^2$.
(b) Violation of the TUR as a function voltage, with exact calculations for ${\cal F}$.
Parameters are $\epsilon_d =0, \beta=1$ and $\Gamma=4$ in (b).
}
\label{Figsd1}
\end{center}
\end{figure}


\subsection{Model A: Single dot junction}
The junction includes a single site of energy $\epsilon_d$ coupled to two leads. The transmission function is given by
\be
{\cal T}(E) = \frac{\Gamma_L \Gamma_R} { (E- \epsilon_d)^2 + (\Gamma_L + \Gamma_R)^2/4},
\label{eq:singletrans}
\ee
where we assume the wide band limit for the spectral density of the baths, 
thus taking $\Gamma$ as energy independent. 

In the weak coupling limit, the transmission function is sharply peaked around $\epsilon_d$, 
\bea
{\cal T}(E) &=& 2 \pi \frac{\Gamma_L \Gamma_R}{\Gamma_L + \Gamma_R} \delta (E-\epsilon_d), \\
{\cal T}^2(E) &=& 4 \pi \frac{\Gamma^2_L \Gamma^2_R}{(\Gamma_L + \Gamma_R)^3} \delta (E-\epsilon_d).
\eea
From here, we readily calculate the integrals ${\cal T}_1$ and ${\cal T}_2$, given as
\bea
{\cal T}_1 &=& \frac{1}{\hbar}\frac{\Gamma_L \Gamma_R}{\Gamma_L + \Gamma_R},
 \nonumber \\
{\cal T}_2 &=& \frac{1}{\hbar}\frac{2 \Gamma_L^2 \Gamma_R^2}{(\Gamma_L + \Gamma_R)^3}.
\eea
Since $(\Gamma_L-\Gamma_R)^2>-\Gamma_L\Gamma_R$, the condition in Eq.~(\ref{eq:cond-trans}) 
cannot be satisfied and therefore the TUR is valid. In particular, when $\Gamma_L=\Gamma_R$,
we find that ${\cal T}_2/{\cal T}_1=1/2$. In fact, this ratio is prevalent: systems with 
a set of sharp resonances show up this ratio at the weak coupling limit. 

In the very strong coupling regime, $\Gamma > \epsilon_d, k_B T$, we 
perform an asymptotic expansion with $\Gamma$,
${\cal T}(E)\sim  1- \frac{\epsilon^2}{\Gamma^2} + \frac{\epsilon^4}{\Gamma^4}$. We plug it into
Eq. (\ref{eq:F}) and get
\be
\beta V \frac{\langle \langle j^2 \rangle \rangle}{\langle j \rangle} = 2 + \frac{\beta^2 V^2}{6} \Big[ \frac{a}{(\beta \Gamma)^2} + \frac{b}{(\beta \Gamma)^4} + \cdots \Big],
\ee
where $a<0$ and $b>0$. 
Specifically, $a=(\pi^2/3-4)$ and $b=\frac{7\pi^4}{15}-18 + \frac{\pi^2}{3}(\frac{\pi^2}{3}-4)$.
Altogether, we find that for a single resonant level the TUR is satisfied at weak coupling 
when indeed the dynamics can be described by a Markovian population dynamics \cite{Prager}.
Nevertheless, it is violated at strong coupling when high-order tunneling processes contribute
and the Markovian population dynamics breaks down \cite{emary}. 
In fact, under the $V^2$ expansion, the TUR is always violated at strong coupling 
in the single dot model, with the function  $C_{\rm neq}$ approaching zero from below.

In Fig. \ref{Figsd1}(a), we display the violation of the TUR in the single dot model
based on the analytical $V^2$ expansion, Eq. (\ref{eq:TURferm}). We further compare simulations to
the exact form, received from exact expressions for the current and noise, Eq. (\ref{eq:levitov}).
We find that the quadratic formula (\ref{eq:F}) very well captures the deviation from linear response up to 
$\beta V\approx1$. 
Fig. \ref{Figsd1}(b) further shows that only at high voltage, $\beta V\gg 1$, when dissipation is excessive,
the TUR is satisfied. 

\begin{figure} [htbp]
\begin{center}
\includegraphics[scale=0.6]{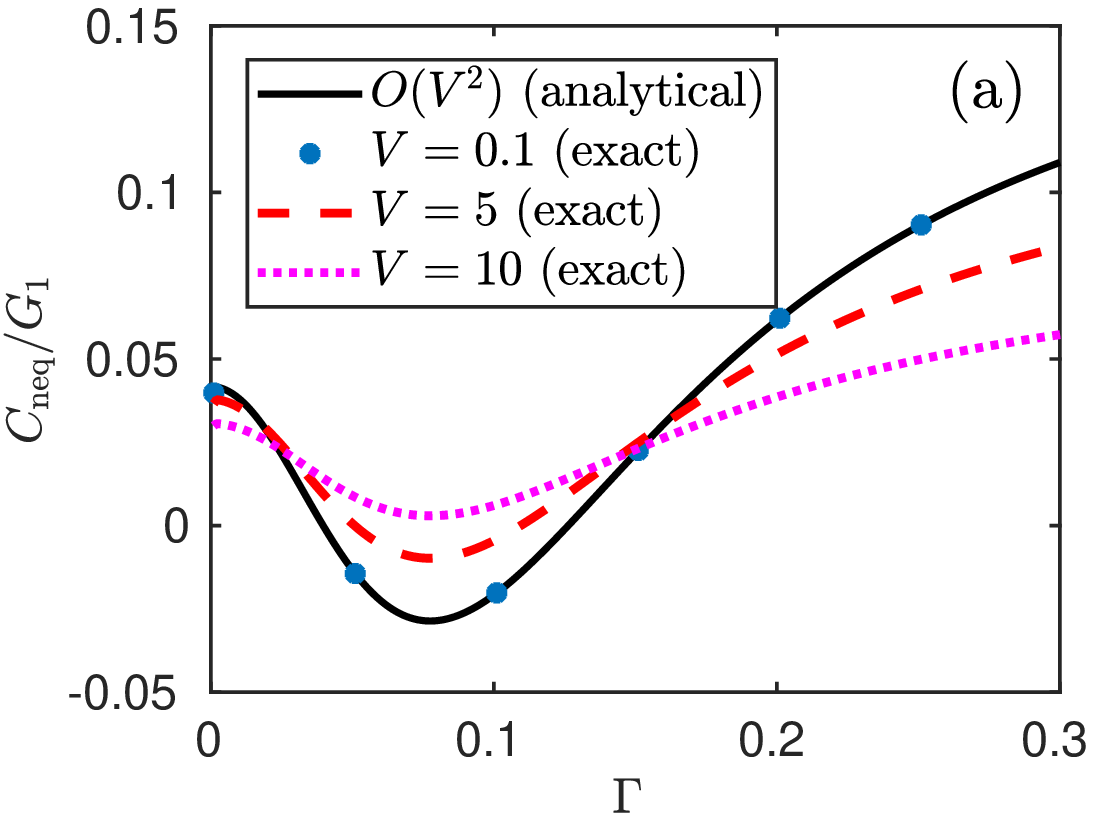}
\includegraphics[scale=0.6]{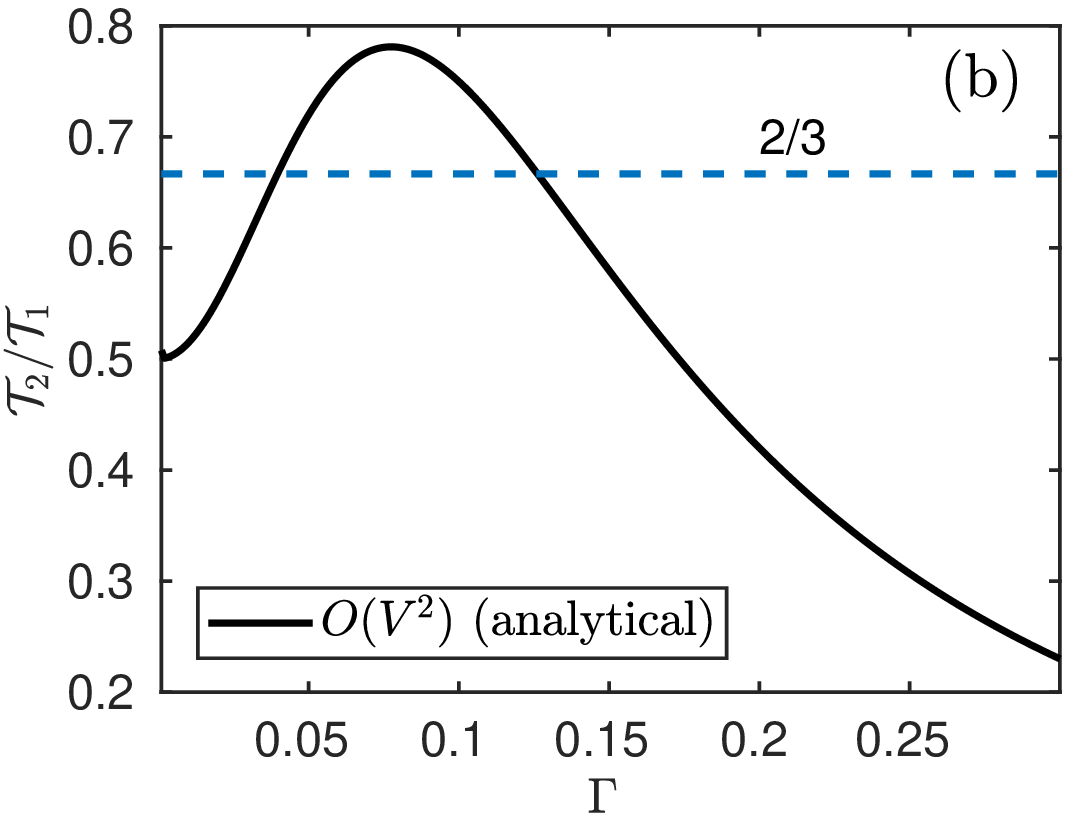} 
\caption{ 
Double quantum dot: (a) Violation of the TUR as a function of $\Gamma$ based on Eq.~(\ref{eq:TURferm})
with the transmission function given in Eq.~(\ref{eq:DDtrans});
exact calculations display ${\cal F}/V^2$.
Here $\epsilon_L=\epsilon_R$=0, $\beta$=1, $\Omega$=0.05, $\Gamma=0.08$.
(b) When ${\cal T}_2/{\cal T}_1>2/3$, 
the TUR is violated at weak coupling.
}
\label{dd1}
\end{center}
\end{figure}
%

\begin{figure} [ht]
\begin{center}
\includegraphics[scale=0.6]{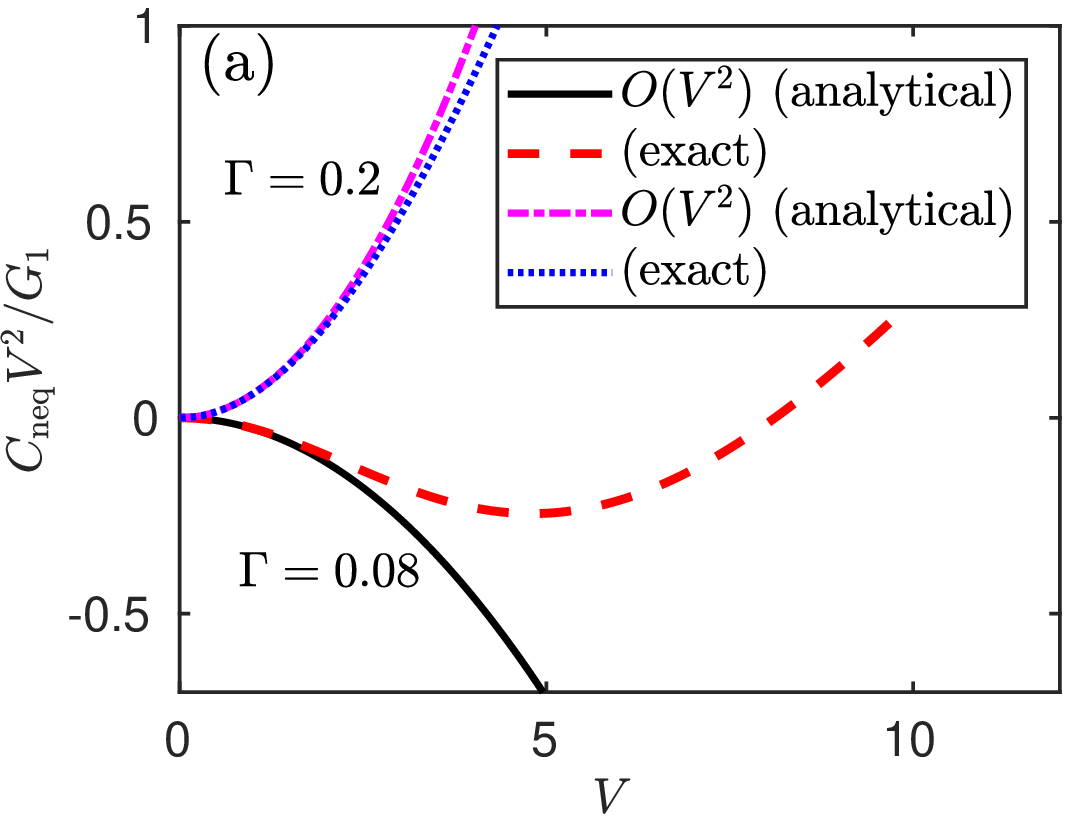}
\includegraphics[scale=0.6]{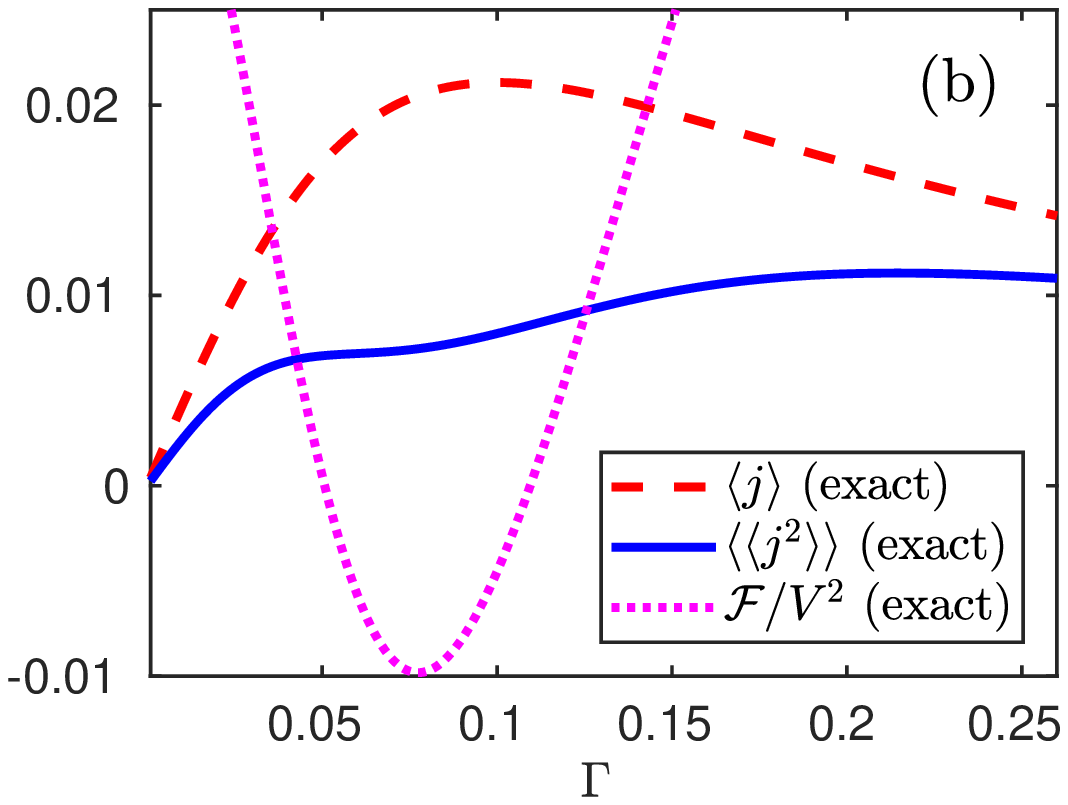}
\caption{
Double quantum dot: (a) Violation of the TUR as a function of bias $V$
based on Eq.~(\ref{eq:TURferm}) with the transmission function given in Eq.~(\ref{eq:DDtrans})
compared to exact calculations  for ${\cal F}$.
Here $\epsilon_L=\epsilon_R=0, \beta=1, \Omega=0.05$.
(b) The exact charge current $\langle j \rangle$, associated noise $\langle \langle j^2 \rangle \rangle$,
and the TUR violation function ${\cal F}/V^2$ plotted against $\Gamma$. 
Here $V=5$. Other parameters are same as in panel (a).}
\label{dd2}
\end{center}
\end{figure}


\subsection{Model B: Serial double dot junction}

The serial double dot junction includes two sites of energies $\epsilon_{L,R}$, 
which are coherently coupled to each other through the tunneling element $\Omega$. 
The dots are individually coupled to the respective leads with
dot $\nu$ coupled to the $\nu-th$ metal with the hybridization energy $\Gamma_{\nu}$.
The transmission function of a double dot is given by \cite{TURcoh,interf4,lena15}
\be
{\cal T}(E)= \frac{\Gamma_L \Gamma_R \Omega^2} {|(E-\epsilon_L + i \Gamma_L/2) (E-\epsilon_R + i \Gamma_R/2) - 
\Omega^2 |^2}.
\label{eq:DDtrans}
\ee
%
%
Assuming levels' degeneracy, $\epsilon_L=\epsilon_R$,
a Markovian master equation neglecting coherences is generally inaccurate and cannot 
describe transport in this model even at weak coupling \cite{Prager,schon,Mukamel,Gernot,flindt}. Thus, we expect TUR violations in this model 
when $\Gamma, \Omega< k_BT$. 
Assuming a symmetric coupling $(\Gamma_L=\Gamma_R=\Gamma)$, 
we perform the integration (\ref{eq:Tn}) and obtain \cite{TURcoh}
\bea
{\cal T}_1 &=& \frac{1}{\hbar}\frac{2 \Gamma \Omega^2}{ \Gamma^2 + 4 \Omega^2},
\nonumber\\
{\cal T}_2 &=&\frac{1}{\hbar} \frac{4 \Gamma \Omega^4 ( 5 \Gamma^2 + 4 \Omega^2) }{ (\Gamma^2 + 4 \Omega^2)^3}.
\label{eq:T12dd}
\eea
According to Eq. (\ref{eq:cond-trans}), a violation of the TUR at weak coupling 
is expected when ${\cal T}_2/{\cal T}_1 >2/3$,
translated here to $4x^2-7x+1<0$  
with $x=\Omega^2/\Gamma^2$. This prediction is quite accurate as we find in Fig. \ref{dd1}.

In contrast, at strong coupling, $\Gamma \gg \Omega,k_B T$, the transmission function is suppressed
 because of a destructive interference effect \cite{interf3,interf4,lena15}. In this limit
the TUR is satisfied; recall that we proved in Eq. (\ref{eq:low-trans}) that the TUR is valid
when the transmission is small.
We can also prove this result analytically. In the very strong coupling regime, 
$\Gamma \gg \Omega, k_B T$, an asymptotic expansion in $\Gamma$ gives
${\cal T}(E)\sim 16 \frac{\Omega^2}{\Gamma^2} + {\mathcal O} \left(\frac{\Omega^4}{\Gamma^4}\right)$.
We plug this expansion into Eq. (\ref{eq:F}) and find that $C_{\rm neq}>0$ to order $1/\Gamma^2$.

We display our results for the double dot model in Figs. \ref{dd1} and \ref{dd2}.
First, in Fig. \ref{dd1}(a) we show that the TUR can be violated at small $\Gamma$, specifically here
in the range $0.04<\Gamma<0.126$. 
This observation agrees with Ref. \cite{TURcoh}. 
We again confirm that Eq.~(\ref{eq:F}) very well performs even at high voltage, 
compared to exact calculations from Eq. \ref{eq:levitov}.
We further verify in panel (b) that TUR violations take place precisely when ${\cal T}_2/{\cal T}_1>2/3$.
This inequality thus serves as an excellent estimate of TUR-breaking at weak coupling. 

We uphold the quadratic approximation (\ref{eq:F}) in Fig. \ref{dd2}(a) by manifesting
that it is valid up to $\beta V\sim 2$, whether or not the TUR is verified.
Finally, in Fig. \ref{dd2}(b) we display the players beyond the TUR: the current and its fluctuations.
The TUR is violated when the current is enhanced but the noise is suppressed, 
which is a favorable regime of operation.


\begin{figure}
\includegraphics[scale=0.6]{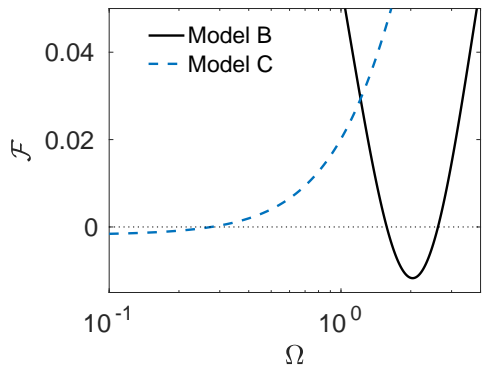}
\caption{
Uncovering TUR violations in model B ($\Omega/\Gamma\sim 1$) and Model C
($\Omega/\Gamma\ll1)$.
We use $\beta=1$, $V=1$ and $\Gamma=4$.}
\label{side}
\end{figure}

\subsection{Model C: Side-coupled double dot junction}

The side-coupled model includes two electronic sites of energies $\epsilon_{1,2}$, where
site 1 is directly coupled to the two metals, and site 2 is side-coupled to site 1.
Assuming the two sites are degenerate at the Fermi 
energy, $\epsilon_{1,2}=0$, 
as we show in Fig. \ref{FigT}, the side-coupled junction displays a node at $E=0$.
The transmission function of this model is given by
\bea
{\cal T}(E) = 
\frac{E^2 \Gamma_L \Gamma_R}{|\big(E+ \frac{i}{2} (\Gamma_L + \Gamma_R)\big)E - \Omega^2|^2}.
\eea
At weak coupling $\Gamma\lesssim \Omega$, the transmission shows a two-peak structure around $\pm \Omega$.
Given the node at $E=0$, at very weak coupling we can approximate the transmission by two separate (close to) 
Lorentzian functions. We then conclude that similarly to the single dot model, we precisely get
${\cal T}_2/{\cal T}_1=1/2$. 
Therefore, according to Eq. (\ref{eq:cond-trans}), TUR violations are not expected at weak coupling 
in the side-coupled model.
In contrast, at large couplings, $\Gamma\gg \Omega$, the transmission behaves similarly to the 
single dot resonant level model A,
therefore we expect to overturn the TUR.
This could be rationalized by the breakdown of a Markovian master equation for population dynamics in this regime,
due to the contribution of high order tunneling processes \cite{Ivana,Xue}.
Figure \ref{side} displays TUR violations in models B and C using the complete expressions (\ref{eq:levitov}).
It is significant to note that the models behave in a strikingly different way at weak and large splitting
$\Omega/\Gamma$.

We summarize our findings on the validity of the TUR for single, serial double and side-coupled double quantum dot setups 
in the Table.
\begin{widetext} 
\label{tab:title}
\begin{center}
Table: Thermodynamic uncertainty relation for charge transport in quantum dot setups\\
\begin{tabularx} 
{\columnwidth}{l l l   } 
\hline
\hline
{hybridization $\Gamma$}  \,\,\,\,\,\, &  {single dot (A) and side-coupled (C) models } \,\,\,\,\,\,\, & {serial double dot model (B)}  \\ 
\hline

{weak} & {valid} & {invalid}    \\
  &             (Markovian master equation for population) & (non-Markovian population dynamics)  \\   \\
{strong }& {invalid} & {valid}    \\
        &          (high-order electron tunneling processes)  & (low transmission function)  \\
\hline
\hline
\end{tabularx}
\end{center}
\end{widetext}
\section{Summary}
\label{Sec-summ}

We investigated the validity of the thermodynamic uncertainty relation in a single-affinity junction beyond linear response.
Based on the nonequilibrium fluctuation symmetry, we derived a relation between the current,
its noise, and the entropy production, given in terms of nonlinear transport coefficients. 
From this relation we received a general condition for validating the TUR,  Eqs. (\ref{eq:TUR})-(\ref{eq:f2}).
We exemplified the analysis on a analytically tractable 
noninteracting fermionic system. We derived a general relation
for validating the TUR, which was given in the language of the transmission function [Eq. (\ref{eq:F})], and tested it
with central charge transport models, with single and double (serial and side-coupled)
quantum dots.

Future work will be focused on the exploration of the TUR in interacting (electron-electron and
electron-phonon) quantum systems based on analytical results for the cumulant generating function
\cite{bijay-segal-fcs}, and on the analysis of parallel relations for heat conducting systems \cite{Saito-Dhar-2007}. 
Furthermore, we plan to study non-reciprocating continuous 
quantum heat machines that rely on quantum coherences for operation, 
e.g. absorption refrigerators \cite{Kos,Kilgour,Novotny},
and search for optimized regimes of operation with suppressed fluctuations and high output power.

\begin{acknowledgments}
DS acknowledges the NSERC discovery grant and the Canada Chair Program. 
BKA gratefully acknowledges the start-up funding from IISER Pune and 
the hospitality of the Department of Chemistry at the University of Toronto.
\end{acknowledgments}

\renewcommand{\theequation}{A\arabic{equation}}
\setcounter{equation}{0}  

\section*{Appendix: Transmission functions}

The transmission function that appears in the Levitov-Lesovik formula, Eq.~(\ref{eq:CGF}), 
can be computed from the Green's function of the system and the self-energies resulting 
from its couplings to the leads, 
\be
{\cal T}(E) = {\rm Tr}\Big[\hat{G}_0^r(E) \hat{\Gamma}_L(E) \hat{G}_0^a(E) \hat{\Gamma}_R(E)\Big].
\ee
Here, $\hat{G}_0^{r,a}(E)$ is the retarded (advanced) Green's function for the system and 
$\hat{\Gamma}_{L,R}$ are the hybridization matrices that include the coupling to the reservoirs (electrodes)
 $L$ and $R$. 

{\it Model A.}
The single dot model, also referred to as the resonant transmission model, includes
a single electronic level with a site energy $\epsilon_d$, which is coupled to two leads. 
In this case, the Green's functions are c-numbers,
${G}_0^r(E)= \left[E-\epsilon_d  + i (\Gamma_L(E) +\Gamma_R(E))/2\right]^{-1}$, 
$G_0^a(E)=[G_0^r(E)]^{\dagger}$. In general, the hybridization parameter depends on energy, $\Gamma(E)$. 
Nevertheless, for simplicity, in the main text we use a wide band model with a fixed value for $\Gamma$.
The transmission function is given by 
\be
{\cal T}(E) = \frac{\Gamma_L(E) \Gamma_R(E)} { (E- \epsilon_d)^2 + (\Gamma_L(E) + \Gamma_R(E))^2/4},
\ee
reducing to Eq.~(\ref{eq:singletrans}) in the wide band limit.

{\it Model B.}
The serial double quantum dot setup comprises two levels with site energies $\epsilon_L$ and $\epsilon_R$. The dots 
are coupled to each other with a coherent tunneling $\Omega$. Each dot is furthermore coupled 
to its respective metal lead. The Green's functions of the system are $2\times 2$ matrices, given by
\bea
\hat{G}_0^r(E)= \big[E \hat{I}-\hat{H}_S- (\hat{\Sigma}^r_L(E)+\hat{\Sigma}^r_R(E))\big]^{-1}, 
\eea
where the system Hamiltonian is
\bea
\hat{H}_S=
\begin{bmatrix}
\epsilon_L & \Omega\\
\Omega & \epsilon_R 
\end{bmatrix} 
\label{eq:HS}
\eea
with $\hat{\Sigma}_L(E) = -\frac{i}{2} \hat{\Gamma}_L(E)$, and similarly for $\hat{\Sigma}_R(E)$.
The self-energy matrices are
\bea
\hat{\Gamma}_L(E)=
\begin{bmatrix} 
\Gamma_L(E) & 0 \\
0 & 0 
\end{bmatrix},
\quad 
\hat{\Gamma}_R(E)=
\begin{bmatrix} 
0 & 0 \\
0 & \Gamma_R(E) 
\end{bmatrix}
\label{eq:GA}
\eea
The resulting transmission function is 
\bea
{\cal T}(E)= 
\frac{\Gamma_L(E) \Gamma_R(E) \Omega^2} 
{|[E-\epsilon_L + i \Gamma_L(E)/2] [E-\epsilon_R + i \Gamma_R(E)/2] -
\Omega^2 |^2}.
\nonumber\\
\eea
%
In the wide band limit we acquire Eq.~(\ref{eq:DDtrans}). 

{\it Model C.}
The side-coupled model includes two dots of energies $\epsilon_{1,2}$, which are coherently coupled.
In this design, level 1 is coupled to both metal leads while level 2 does not directly connect to the electrodes.
The system Hamiltonian $\hat H_S$ and the matrix $\hat{\Gamma}_L(E)$ are 
given by Eqs. (\ref{eq:HS}) and  (\ref{eq:GA}). Since level 1 couples to both leads,
$\hat{\Gamma}_R(E)$ is given by
\bea
\hat{\Gamma}_R(E)=
\begin{bmatrix} 
\Gamma_R(E) & 0 \\
0 & 0
\end{bmatrix}
\eea
Gathering the transmission function we get
\be
{\cal T}(E) 
= \frac{(E-\epsilon_2)^2 \Gamma_L(E) \Gamma_R(E)}
{\Big|\big[E-\epsilon_1 + \frac{i}{2} (\Gamma_L(E) + \Gamma_R(E))\big] \big(E-\epsilon_2\big) - \Omega^2\Big|^2}.
\ee
Using $\epsilon_{1,2}=0$, we find that the transmission shows a node at $E=0$, and
a double-peak structure at $E=\pm \Omega$. 


\end{document}